\documentclass{iaus}
\usepackage{graphicx}

\newcommand{\pc}{{\rm pc}}
\newcommand{\msun}{{M_\odot}}
\newcommand{\Msun}{{M_\odot}}
\newcommand{\Lsun}{{L_\odot}}

\newcommand{\kms}{{\rm km}{\rm s}^{-1}}
\newcommand{\citep}{\cite}
\newcommand{\beq}{\begin{equation}}
\newcommand{\beqa}{\begin{eqnarray}}
\newcommand{\eeq}{\end{equation}}
\newcommand{\eeqa}{\end{eqnarray}}

\title{How to form bulges/ellipticals in dark halos as fast as central black holes?} 
\author{HongSheng Zhao$^1$, Bing-Xiao Xu$^2$, and Xue-Bing Wu$^2$}
\affiliation{
$^1$SUPA, University of St Andrews, KY16 9SS, Fife, UK \\ [\affilskip]
$^2$Department of Astronomy, Peking University, 100871
Beijing, China,\\Email: 
hz4@st-andrews.ac.uk, xubx@bac.pku.edu.cn, wuxb@bac.pku.edu.cn}

\pubyear{2007}
\volume{xxx}  
\pagerange{001--004}
\date{17 July 2007 and in revised form \today}
\jname{Proceedings 245 IAU Symposium}
\editors{Martin Bureau}
\begin{document}

\maketitle

\begin{abstract}
Gravity is nearly a universal constant in the cusp of an NFW galaxy halo.   
Inside this external field an isothermal gas sphere will collapse 
and trigger a starburst if above a critical central pressure.  
Thus formed spheroidal stellar systems have Sersic-profile and satisfy the 
Faber-Jackson relation.  The process is consistent with observed starbursts.
We also recover the $M_{BH}-\sigma_*$ relation, if the gas collapse 
is regulated or resisted by the feedback from radiation from the central BH.
\end{abstract}



\section{Tight correlation of formation of Black Hole and Bulges}

The formation of central black holes (BHs) in galaxies is likely a rapid process 
since most quasars have already formed at redshift $z>2$.  
The co-relation between the BH mass and 
the velocity dispersion of the spheroidal (bulge) component 
is so tight that it is hard to explain 
unless bulges form as fast as BHs to keep their growth neck-to-neck.  
While at the present day the BH accretion rate is completely 
decoupled from the bulge growth, 
it is possible that their growth was correlated during the violent feedbacks 
at high redshift.  Indeed starburst activities peak at similar redshifts as 
the quasars as a whole.  

In two recent papers by Xu, Wu, \& Zhao (2007) and Xu \& Wu (2007), we propose that 
bulges can form by a rapid collapse due to radial instability of isothermal gas.  
This model has the nice feature of forming bulges before disks.

Here we iterate the key steps of the above scenario, 
but without invoking the gravothermal instability as in Xu et al. (2007).  
Instead we follow the example of Elemgreen (1999) 
and find the equilibrium configurations of 
the maximum gas mass  inside an external gravity.  
We also generalize the argument to a star-gas mixture to show 
that we can form bulges with a reasonable profile.
Assuming the rapid star bursts in the bulge are regulated by the accretion-driven 
wind of the central black hole, we derive the BH mass-stellar dispersion relation.

\section{A universal constant gravity scale for dark halos}

In the Cold Dark Matter (CDM) framework, 
baryons fall into the potential well of CDM, cool and condense
into stars.  Here we consider the properties of gas and 
stellar equilibrium in the external field of dark matter.

The background dark matter distribution is often described
by the NFW density distribution for dark matter (Navarro, Frenk \& White 1997), which has a density 
$\rho_{NFW} \approx \rho_s r_s/r$ inside a scale radius $r_s$. 
In the central region which concerns the galaxy bulge, we note 
{\it an interesting universal scale for the dark halo gravity}
\beq 
g_{\rm DM}(r)=\frac{GM_{DM}(r)}{r^2} = 2 \pi
G \Pi \sim 10^{-10}{\rm m}\ {\rm sec}^{-2} \Xi, \qquad \Xi \sim 1
\eeq
where 
$\Pi=\rho_s r_s \sim  130 M_{\odot} {\rm pc}^{-2} \Xi$ is a column density
and $\Xi(M_{vir},z,c) \sim 1$ 
is a shallow function of the halo virial mass
$M_{vir}$, the redshift, and the concentration $c$.
In another words, the gas where formed the bulge was imbedded in an
uniform external field from the dark matter potential.  We can also define 
a dark matter central pressure
\beq
P_{DM} \equiv g_{DM}^2/(4\pi G)=g_{DM}\Pi/2 = \rho_{DM}(r) \cdot g_{DM} \cdot (r/2)
\eeq
for later use.

\section{Maximum gas mass sustainable by halo gravity}

In general for a gas and a stellar sphere 
imbedded in an external DM gravity,
Consider imbedded in an external DM gravity
an isothermal gas sphere $\rho(r)$ of sound speed $\sigma$,
and an isotropic stellar sphere $\rho_*(r)$ of dispersion
$\sigma_*$ in quasi-static equilibrium, 
the potential at a given radius $r$ is 
\beq\label{Phi}
\Phi = \int_0^r \left( g_{DM}+{GM+G M_* \over r^2}\right) dr.
\eeq
The equilibrium satisfies the equations
\beqa
\left( g_{DM}+{ G M + GM_* \over r^2}\right) 
&=& -{\sigma_1^2 d\ln[\sigma^2 \rho(r)] \over dr} 
= -\sigma_*^2{d\ln[\sigma_*^2\rho_*(r)] \over dr},\\
4\pi r^2 &=& {dM(r) \over \rho(r) dr} = {dM_*(r) \over \rho_*(r) dr},
\eeqa 
where we define $\sigma_1^2 \equiv (1+\Gamma) \sigma^2$.
Here a position-independent feedback factor $\Gamma \gg 1$ is introduced because  
the radiation from the star burst and the accreting central black hole can
generate an additional opacity-induced pressure $(\rho \sigma^2)\Gamma$ on the dusty gas sphere 
(but not the stellar sphere), countering the gravity.

First consider the stage where the stellar mass is negligible, so $M_* \ll M$.  
Rewrite the equations in term of the following dimensionless mass, radius and density, 
 \beq
 m \equiv {M \over \sigma_1^4 g_{DM}/G},
 \qquad x(m) \equiv {r \over \sigma_1^2/g_{DM}},\qquad
 p(m)={\rho(M) \sigma_1^2 \over g_{DM}^2 (4\pi G)^{-1} },
\eeq
and express the rescaled gas mass $m$ as the independent coordinate, 
the problem is recasted to solving the pair of dimensionless ODEs
\beq\label{ode}
-{x^2 dp(m) \over dm} = 1 + {m \over x(m)^2}, \qquad
 {x^2 dx(m) \over dm} = {1 \over p(m)} .
\eeq
For each value of $p(0)$, the gas density profile under the
hydrostatic equilibrium can be totally determined with the following
initial conditions at the center for the radius $x(0)=0$ and the rescaled density
\beq
p_0 = p(0) = {\rho_0\sigma_1^2 \over P_{DM} }, \qquad P_{DM} \equiv {g_{DM}^2 \over 4\pi G }, 
\eeq
where $p(0)/(1+\Gamma)$ equals the ratio of
the gas central pressure $\rho_0\sigma^2$ vs the dark matter's 
pressure under self-gravity $P_{DM}$.

\begin{figure}
\includegraphics[height=7cm, width=7cm,angle=0]{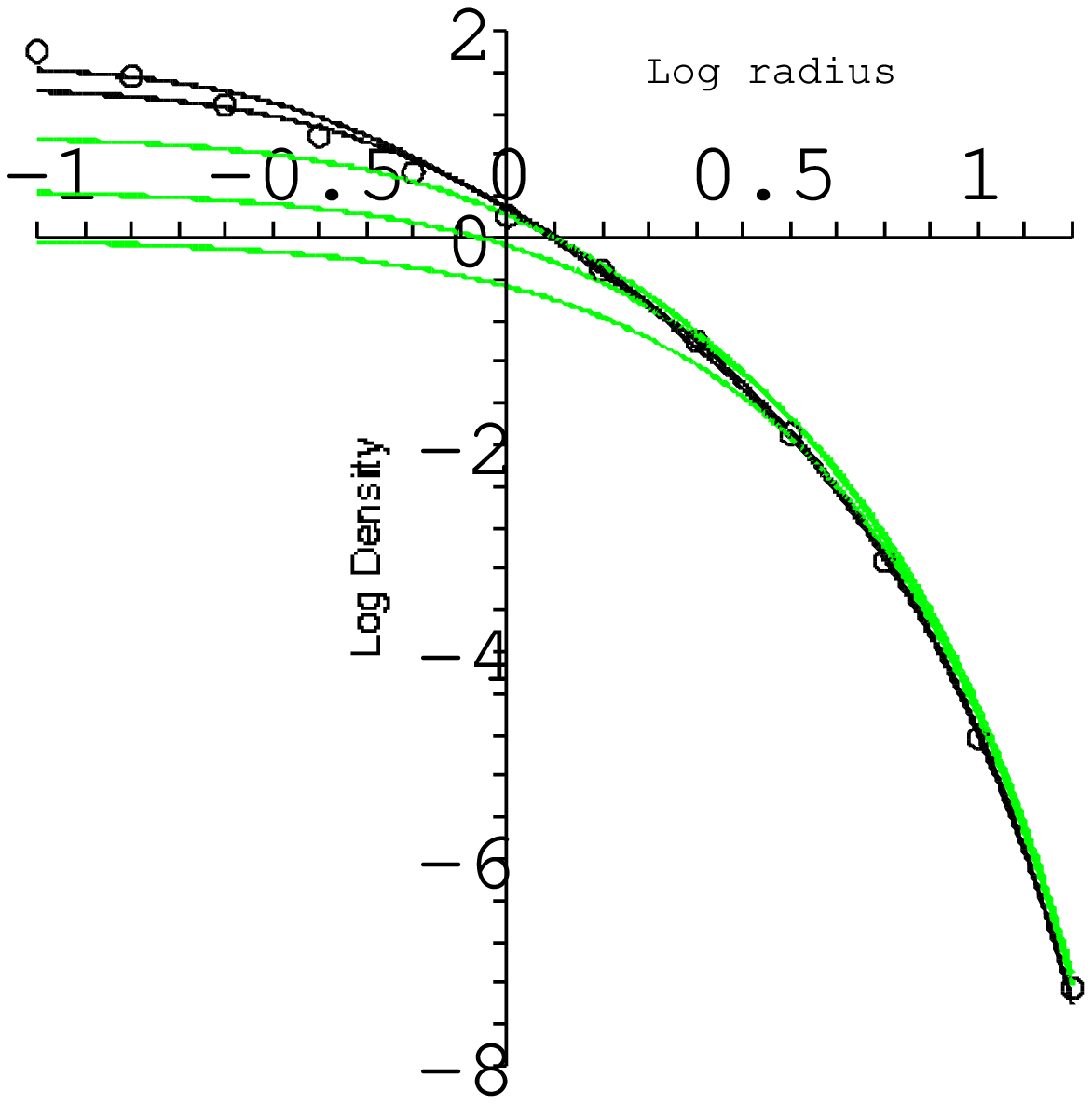} 
\includegraphics[height=7cm, width=7cm,angle=0]{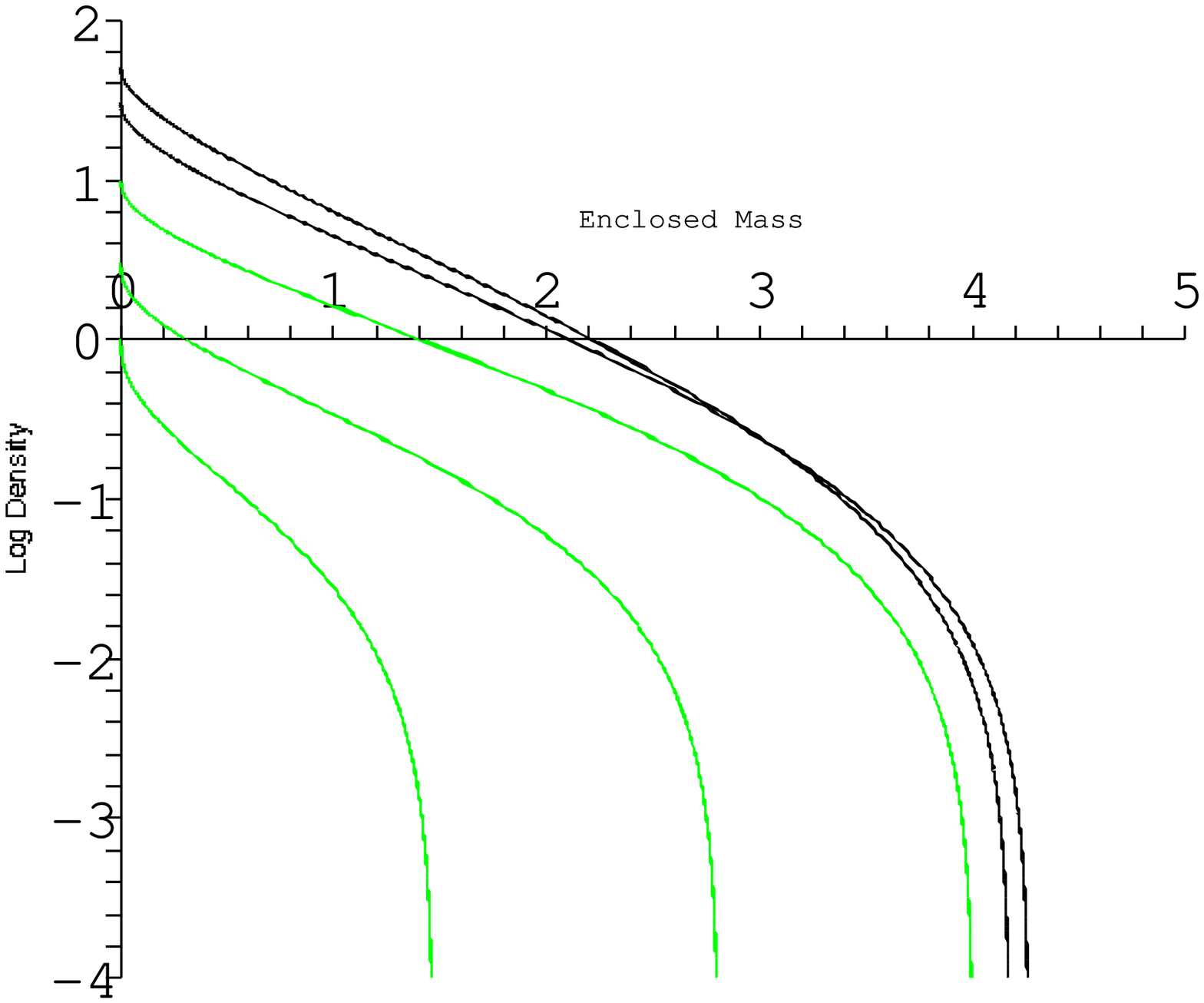} 
\caption{shows mass distribution of 
an isothermal gas sphere embedded in a NFW dark matter cusp of a uniform gravity $g_{DM} \sim 10^{-10}$m/sec$^2$ for  models (from bottom to top)
with increasing dimensionless central gas pressure
$p_0=1,3,10$ (in solid green), and $p_0=30,100$ (black dashed).
Panel (a) 
shows the models in $\log p(m)$ vs $\log x(m)$ (rescaled density vs. rescaled radius $x(m)={r \over \sigma^2/g_{DM}}$).  
Note how the black curves are above the green curves at small radii, but 
dip below the green curves at large radii, a feature of reaching a maximum in total gas mass at the critical pressure ($p_0=30$).  
Also shown is a Sersic $(n=1.2)$ profile (red circles).
Panel (b) 
shows $\ln p(m)$ (the logarithm of rescaled gas density or pressure, 
$p(m)={\rho(M) \sigma^2 \over g_{DM}^2/(4 \pi G)}$)
as function of the rescaled 
enclosed gas mass $m=M/(\sigma^4/G/g_{DM})$, cf. eq. 3.4). 
Note the total gas mass $m$ increases with $p_0$ until the critical 
value $p_0 \sim 30$, afterwards the mass decreases with central pressure.}\label{fig:gasden}
\end{figure}

Computing the gas equilibrium for a range of core pressure $p_0$
(see figure~\ref{fig:gasden}), we find that the gas density generally 
falls montonically with radius or mass.  All models have finite
mass out to infinite radius where $\rho=0$.  This interesting
behavior is due to the deep potential well of the external gravity,  
which makes the isothermal density drop exponentially with radius
as 
$-{ \sigma_1^2 \over g_{DM}} \ln {\rho(M) \over \rho_0} \sim r(M),$ 
hence the mass converges quickly if neglecting self-gravity.

The finite mass of these gas spheres will give another interesting 
behavior.  There is a critical core pressure
\beq p_0 = {\rho_0 \sigma_1^2 \over P_{DM}} \approx 30 
\eeq 
above which the gas density $\rho(M)$ of a parcell of gas $dM$ 
no longer increases monotonically with an increase of central pressure,
and in fact the total mass will decrease with increasing 
$p_0$ after it reaches a maximum value 
\beq 
M_{max} \approx 4.3  \left({\sigma_1^4 \over g_{DM} G}\right).
\eeq

These limits on gas central pressure and total mass are related 
to the instability first discussed by Elmgreen (1999).
Gas sphere above certain critical mass $M_{max} \propto \sigma_1^4$ 
or critical central gas density or pressure 
do not have stable solutions; adding tiny amount of gas would lead to collapse.  
It is interesting to speculate that the bulge
formation originates from such a gas instability.    

\section{Post-starburst Mass Profile} 

Our models are also {\it generalizable} while gas is converting to stars.
The simplest solution of eqs. (3.2-3.3) would be a model 
where {\it stars trace gas} radial distribution, so we have 
\beq
{\rho_*(r) \over \rho(r)}={M_*(r) \over M(r)} = {f_* \over 1-f_*}, \qquad
{\sigma_*^2 \over \sigma^2} = 1+\Gamma \gg 1,
\eeq
where the position-independent factor 
$f_*(t)$ is the fraction of gas formed into stars at time $t$.  Such a solution is 
possible if the feedback is regulated by star formation.
Rescaling the gas-only solution (eq. 3.7), we obtain the stability criterion
\beq
{\rho_0 \sigma^2 \over (1-f_*)/(1+\Gamma) } 
= {\rho_{*,0} \sigma_*^2 \over f_* }
\le 30 P_{DM}, \qquad P_{DM}={g_{DM}^2 \over 4\pi G}.
\eeq

Gas can turn into stars quasi-staticly where maintaining the above equality
at the critical density and mass.  Such formed stellar system would have a total mass
\beq
M_{*}^\infty = {4.3 \sigma_*^4 \over g_{DM} G } 
\sim 4\times 10^{11} \Msun \left({\sigma_* \over 200\kms}\right)^4, 
\eeq
and a profile
\beq
{M_*(r) \over M_*^\infty f_*(t)} = \int_0^y j(y) (4 \pi y^2 dy), 
\qquad y={r \over \sigma_*^2 g_{DM}^{-1}}, \qquad
j(y) \approx 0.64 y^{-1} \exp(-1.6 y^{1/1.2}),
\eeq
where $j(y)$ is numerically fitted by a Sersic profile 
(of the volume density with a total mass unity).  
Integrating the density, we find the central surface density 
$I_*(0) f_*^{-1} \sim 2 g_{DM}/G \sim 10 \Pi \sim 1300 \msun\pc^{-2}$
for these systems, independent of the initial gas dispersion $\sigma$ 
and the feedback parameter $\Gamma$.  Eventually $f_*=1$ 
when the gas is exhausted by star formation, and 
we form a stellar system with a bulge-like density profile.
Our model resembles real bulges in term of the central brightness $I_*(0)$
and the Faber-Jackson-like relation $M_*^\infty \sim \sigma_*^4$
between the total mass and stellar dispersion.

During the star burst (SB), the central Black Hole (BH) accretes.   
the SB and BH emit photons with a luminosity $L_{SB+BH}$, 
which diffuse out of the gas sphere while keeping the gas isothermal.
The momentum deposite rate ${2 L_{SB+BH} \over c}$ of photons 
drives an overall feedback force acting on the gas, which can be computed by
\beq
{2 L_{SB+BH} \over c} 
= F(t) 
= \int_0^\infty d(4 \pi r^2) (\rho\sigma^2) \Gamma 
\sim {10 (1-f_*) \Gamma \over 1+\Gamma} {\sigma_*^4 \over G} 
\sim 2 (1-f_*) M_*^\infty g_{DM}
\eeq
insensitive to $\Gamma$ if $\Gamma \gg 1$.  The luminosity and force 
are maximum initially and die out as the star burst finishes.  
If the maximum luminosity at the onset of star burst is contributed fifty-fifty 
between the luminosity of the SB and the Eddington luminosity of the BH, then we obtain 
\beq
{M_{BH} \over 10^8\Msun} = { L_{SB} \over 10^{13}\Lsun } 
= \left({\sigma_* \over 200\kms}\right)^4,
\eeq
which agree with the observed scaling relations 
of the stellar dispersion $\sigma_*$ with the BH mass and
the star burst luminosity respectively.  Assuming the usual SB efficiency of 0.001, 
the star formation time scale is $0.001 M_*^\infty c^2/L_{SB} \sim 0.04 H_0^{-1}$,
comparable to the free-fall time scale $\sim 0.1$Gyr.  The short time scales and  
high luminosity are consistent with the assumption of violent feedback $\Gamma \gg 1$.



\end{document}